\documentclass[journal]{IEEEtran}
\usepackage[style=ieee,backend=biber,sorting=none,maxnames=6,minnames=1, url=false]{biblatex}
\usepackage{amsmath}
\usepackage{enumitem}
\usepackage{amssymb}
\usepackage{graphicx}
\usepackage{soul}
\usepackage{xcolor}
\usepackage{pifont} 
\usepackage{hyperref}  

\AtEveryBibitem{%
  \clearfield{note}%
  \clearfield{addendum}%
  \clearfield{urldate}%
  \clearfield{issn}%
}

\ifCLASSINFOpdf

\begin{document}

\title{Momentum-Based Access and Speed Control for Improved Safety in Heterogeneous Road Networks}

\author{Felix~Wieberneit,
        Emanuele~Crisostomi,
        Wynita Griggs,
        and~Robert~Shorten
\thanks{F. Wieberneit (corresponding author email: fw1520@ic.ac.uk) and R. Shorten are with the Department of Design Engineering, Imperial College London, UK. E. Crisostomi is with the Department of Energy, Systems, Territory and Constructions Engineering, University of Pisa, Italy. W. Griggs is with the Department of Civil and Environmental Engineering, and the Department of Electrical and Computer Systems Engineering, Monash University, Clayton, Victoria, 3800, Australia.}

\thanks{This manuscript has been submitted to the International Journal of Control.}
}
\markboth{}%
{}

\maketitle
\begin{abstract}
The increasing variety of means of transportation, including light vehicles like e-scooters and e-bikes, together with the increasing weight of conventional vehicles due to electrification and consumer preferences for SUVs, are raising serious concerns regarding the safety of road networks. In this paper we design a two-level control algorithm to improve the safety of heterogeneous networks: first, an access control strategy decreases the heterogeneity of the network depending on actual traffic conditions; then, a speed control strategy mitigates the probability of serious injuries in potential collisions. Both control strategies are designed based on momentum considerations, as this is regarded as the most influential variable to assess injury risk. The road network mobility simulator SUMO is adopted to implement and validate our proposed control strategies.
\end{abstract}


\IEEEpeerreviewmaketitle

\section{Introduction}

\subsection{Motivation}
Heterogeneity in road networks is significantly increasing nowadays due to new means of transportation, such as electric bicycles, electric scooters, and lightweight micro-cars \cite{oeschger2020}. While the popularity of such micro-mobility solutions is surging, drivers of such new means of transportation, as well as those of other light-weight vehicle categories, are exposed to ever-increasing safety issues, as vehicles on the roads are increasing their weight in general. In fact, there is an increasing penetration of Electric Vehicles (EVs), that are heavier than their combustion-engine counterparts, as they need to accommodate for the mass of the energy storage unit. In addition, light trucks and sports-utility-vehicles (SUVs) are also increasingly popular vehicles choices \cite{li_traffic_2012}, collectively resulting in a large number of heavier vehicles in road networks. Consequently, nowadays it is possible to observe a significant heterogeneity in the mass of vehicles in road networks, and this raises significant concerns regarding the safety of drivers along the roads.

There is a rich literature that has tackled the problem of formally predicting the injury risk of drivers in mixed-mass traffic. For instance, Evans et al. \cite{evans_mass_1993} first showed empirically the power-law relationship between the fatality risk ratio and mass ratio in a collision between two vehicles of unequal mass. Their result was later theoretically explained using fundamental Newtonian mechanics \cite{evans_driver_1994}. Both contributions highlight the same key and somewhat obvious result, which is, in collisions between two vehicles of unequal mass, the lighter car is exposed to a significantly higher crash severity. These theoretical findings have been confirmed by recent analyses. For instance, a recent article \cite{economist2024} shows that the \textit{``heaviest 
1\% of vehicles in the considered dataset, with weight around 6,800lb, suffer
4.1 ``own-car deaths" per 10,000 crashes, on average, compared with around 6.6 for
cars in the middle of the sample weighing 3,500lb, and 15.8 for the lightest 1\% of
vehicles weighing just 2,300lb. But heavy cars are also far more dangerous to other
drivers. The heaviest vehicles in the data were responsible for 37 ``partner-car
deaths” per 10,000 crashes, on average, compared with 5.7 for median-weight cars
and 2.6 for the lightest cars"}. Indeed, even the added weight of an extra passenger can reduce a driver's fatality risk in a frontal collision by 7.5\%, while increasing the risk to the driver of the other constant weight vehicle  by 8.1\% \cite{noauthor_causal_2001}. The vehicle weight externality, i.e., the increased hazard posed to other road users by heavier cars which are safer for their own occupants, has been discussed and quantified at \$ 136 billion annually (US) in \cite{NBERw17170}. Further research highlights the ``arms race" dynamic of vehicle weight choice \cite{li_traffic_2012, white_arms_2004} and estimates that for each life saved by driving a large vehicle, 4.3 additional fatal crashes involving other vehicles occur \cite{white_arms_2004}, with pedestrians being particularly exposed to the greater hazard posed by heavier vehicles \cite{tyndall_pedestrian_2021}.

\subsection{Momentum-based Metrics for Injury Risk in Vehicular Collisions}
\label{Momentum_metrics}

\begin{figure}
    \centering
    \includegraphics[width=\linewidth]{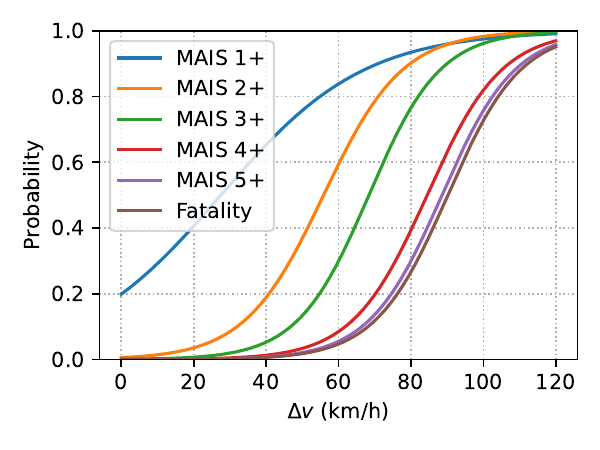}
    \caption{Frontal-collision injury risk as a function of $\Delta v$ (km/h): predicted probabilities of MAIS 1+, 2+, 3+, 4+, 5+, and fatality from univariate logistic models. Coefficients follow the NHTSA curves reported by Wang et al. (2022) \cite{wang_mais0508_2022}.} 
    \label{fig:injury_probability_frontal}
\end{figure}

\begin{figure}
    \centering
    \includegraphics[width=\linewidth]{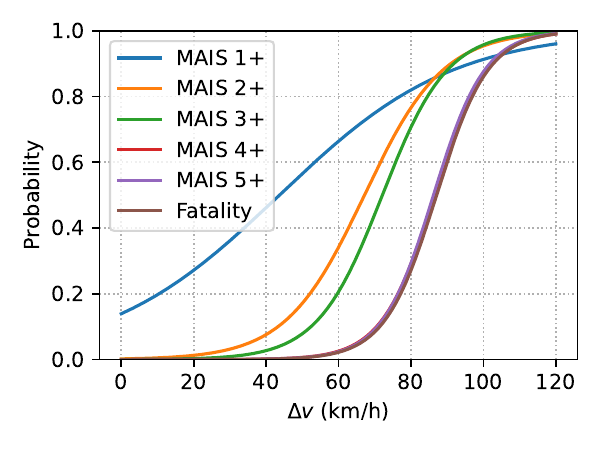}
    \caption{Rear-end collision injury risk as a function of $\Delta v$ (km/h): predicted probabilities of MAIS 1+, 2+, 3+, 4+, 5+, and fatality from univariate logistic models.     Coefficients follow the NHTSA curves reported by Wang et al. (2022) \cite{wang_mais0508_2022}. The lower probability of MAIS 1+ (blue) results from the modelling choice in \cite{wang_mais0508_2022}, and we note here that it is poorly consistent with the curves of more severe injuries for $\Delta v>85$ km/h. In this work, we consider this curve valid for $\Delta v\leq85$.}
    \label{fig:injury_probability_rear-end}
\end{figure}

A widely used collision severity metric (see \cite{evans_driver_1994}, \cite{ sobhani_kinetic_2011}), which embeds the impact of vehicle mass, is $\Delta v$ ---the absolute speed change a vehicle experiences as a result of a collision. The metric is considered as the best available measure of crash severity in the absence of sophisticated crash test equipment \cite{evans_driver_1994}, and can be estimated a priori using the law of conservation of momentum. The relationship between $\Delta v$ and crash injury outcomes is well established: for example \cite{evans_driver_1994} and the US National Highway Traffic Safety Administration publish univariate logistic functions of $\Delta v$ to predict injury outcomes of collisions on the Maximum Abbreviated Injury Scale (MAIS) \cite{wang_mais0508_2022}. MAIS is a medical classification system that categorizes the severity of injuries, ranging from minor (1) to life-threatening (5). MAIS and fatality curves are shown in Figures \ref{fig:injury_probability_frontal} and \ref{fig:injury_probability_rear-end}, for frontal and rear-end collisions, respectively, where category `1+', for instance, refers to an injury of a category greater or equal than 1.

\subsection{Contribution}

As we have explained in Section \ref{Momentum_metrics}, $\Delta v$ has a strong theoretical and empirically demonstrated influence in injury risk, but it is rarely used to dynamically manage traffic risk in practical situations, and only speeds are used as a proxy (i.e., through speed limit signals). The main reason is that $\Delta v$ --- which we remind is the change of speed a vehicle experiences as a result of a collision --- depends on the masses of the two vehicles involved in the accident (as we shall also derive in detail in Section \ref{Momentum_Introduction}), and this information may not be simply available and usable in road networks. Conversely, in this manuscript we shall assume that this information may be actually available in road networks --- leveraging on the recent advances in vehicular communication networks \cite{bhoi2014a, dimitrakopoulos2010, perez2010} --- and we use it within a novel framework for improved safety of vehicles in such heterogeneous vehicular networks. 

In particular, we propose a two-level safety framework to protect a given area of interest where traffic flows of vehicles with different masses co-exist:

\begin{enumerate}
    \item Level 1 [Access control]: We provide --- or not provide --- access to an area based on the mass of the vehicle. Motivated by the asymmetric distribution of injury risk in collisions between vehicles of different masses, we design an ergodic area access control algorithm to shape the vehicular mass distribution on the area of interest. For this purpose, a feedback controller tracks the actual vehicular flow, compares it with the desired distribution, and controls vehicle-class-specific admission probabilities, shaping the ergodic access distribution in favour of lighter vehicles while ensuring stochastic fairness and provable convergence \cite{fioravanti_ergodic_2019}.
    \item Level 2 [Speed control]: We propose a closed-loop area speed-advisory-system that maximizes traffic throughput subject to separate $\Delta v$ constraints on ``risk to self" and ``risk to others" in potential rear-end collisions, for vehicles that have been admitted within the area. The bounds translate to a simple mass-ratio weighted cap on the recommended speed, yielding a one-shot feasible set maximizer for each vehicle. Simulated overtake scenarios highlight how mass mismatch tightens the feasible recommended speed, and show that similar-mass vehicles are more convenient, thus motivating the need of the first level of access-control policy.
\end{enumerate}

This paper is organized as follows: Section \ref{sec:Access_Control} is dedicated to illustrate the first level of safety, i.e., the access control strategy, while Section \ref{sec:Speed_Control} describes how the speed advisory system works for vehicles admitted within the area of interest. Section \ref{Final_Simulation} is used to validate the combined momentum-based access and speed control system, showing an improved safety and efficiency with respect to a single scheme, and especially with respect to an uncontrolled situation. Finally, Section \ref{Conclusions} summarises our work and outlines possible research lines to extend our findings.

\section{Momentum-based Access Control}
\label{sec:Access_Control}

\begin{figure}[h]
    \centering    \includegraphics[width=\columnwidth]{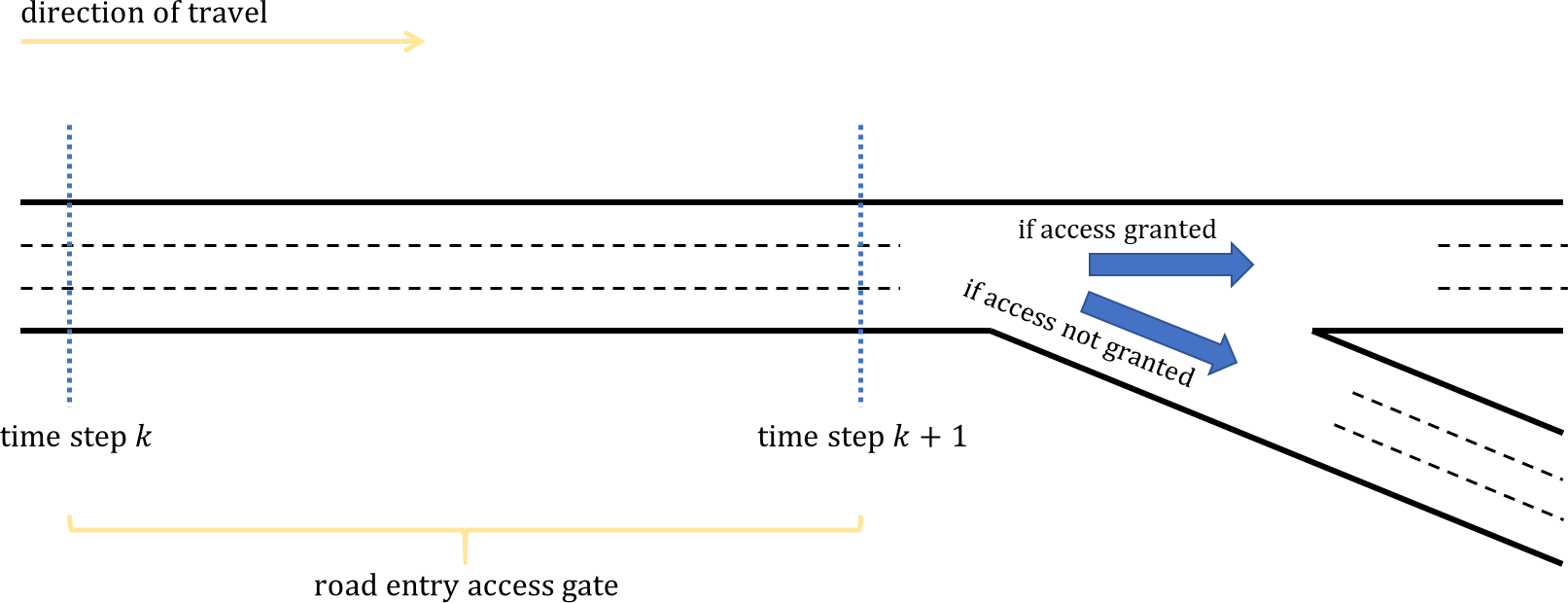}
    \caption{Access control to a restricted area.}
    \label{fig:road_access_scenario}
\end{figure}

The access control problem is exemplified in Figure \ref{fig:road_access_scenario}: a vehicle approaches an area with controlled access, and requests the permission to enter it. If access is denied, the vehicle has to select an alternative road.

For instance, the restricted area may be a city centre, or a sensitive area such as near a school or a health centre, and the objective may be to prioritize light vehicles (e.g., e-scooters or e-bikes). Heavy vehicles may be interested in requesting access to the restricted area (e.g., to avoid taking longer routes around the city), and are usually denied to do so to reduce the probability of collisions with lighter vehicles. However, heavy vehicles may be granted access in special situations (e.g., at night time, when few or no light vehicles are in the restricted area).

In particular, the $i$th vehicle asking for access may be accepted or not, based on its maximum driving momentum along the road $\rho^i_{max}$, which is either dictated by the specific road (i.e., its specific speed limit), or by the class of vehicle (e.g., if its maximum allowed speed is lower than the speed limit of the road). This is formally written as: 
 \begin{gather}
    \rho^i_{max} = \min\{\rho^i_{self}, \rho^i_{road}\},
    \label{rho_max}
\end{gather}
where $\rho^i_{self}$ denotes the momentum of the vehicle calculated according to its mass and the maximum speed that the vehicle is physically capable of travelling at, and $\rho^i_{road}$, which denotes the momentum of the vehicle calculated according to its mass and the maximum speed limit on the road of interest.

\subsection{Mathematical Formulation of the Access Control Methodology}

For simplicity, we now consider a constant flow of vehicles that approaches the restricted area of interest. This flow is equal to the sum of (i) the flow of vehicles admitted in the restricted area, and (ii) the flow of vehicles that have not been given access, and take an alternative road (see Figure \ref{fig:road_access_scenario}).

\begin{figure}[h]
    \centering
\includegraphics[width=\columnwidth]{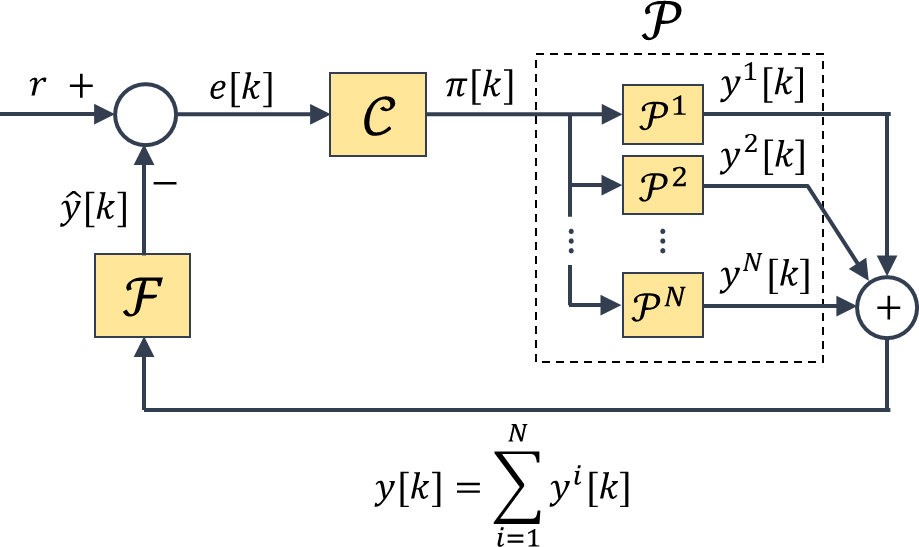}
    \caption{Block diagram of the control strategy in the access control system.}
    \label{fig:interconnection}
\end{figure}

Our proposed access control framework is then schematically illustrated in Figure \ref{fig:interconnection}. In the feedback loop of Figure \ref{fig:interconnection}:
\begin{itemize}
    \item $N$ represents the constant number of vehicles per unit of time that asks for access to the restricted area;
    \item $r$ denotes the desired constant flow of vehicles per time step (e.g., 1 vehicle per minute) that should be granted access to the restricted area;
    \item {$\hat{y}[k]$ denotes the measured flow of vehicles on our road of interest, at time step $k$}, and could correspond to the moving average of the actual vehicles $y[k]$ (e.g., due to measurement delays);
    \item $e[k] := r - \hat{y}[k]$ is the resulting error signal;
    \item $\mathcal{C}$ is any stable, linear shift-invariant, single-input, single-output controller (e.g., a lag controller), and $\pi[k]$ is its output signal;
    \item $\mathcal{F}$ is any stable, linear shift-invariant, single-input, single-output filter (e.g., a pure delay of one time step, or a finite impulse response (FIR) filter);
    \item {$\mathcal{P}^{i}$ denotes the dynamics of the $i$th vehicle};
    \item {and $y^{i}[k] = x^i[k]$ denotes the internal state of the $i$th vehicle, and it could be 1 or 0 depending on whether it is admitted in the restricted area or not}.
\end{itemize}

\subsection{Algorithm Description}
\label{Algorithm_Description}

\begin{enumerate}
    \item At each time step $k$, the controller adjusts its output $\pi[k]$ with the final objective of driving the error $e[k]$ from the reference signal equal to zero;
    \item The output $\pi[k]$ is used to determine the probability of a vehicle that has entered the \emph{road access entry gate} at time step $k$ to be admitted into the restricted area at time step $k+1$. Specifically, we compute the probability $p^i_{\text{Yes}}(\pi[k])$ that the $i$th vehicle is granted access to the restricted area according to a logistic function as
    \begin{equation*}
        p^i_{\text{Yes}}(\pi[k]) = \delta_{l}^i + \frac{\delta_{u}^i}{1+e^{-\lambda^i(\pi[k]-\pi^i_{0})}}, 
    \end{equation*}
    where $\lambda$  denotes the logistic growth rate (i.e., the steepness of the function), $\pi_{0}$ denotes the $\pi$-value of the function's midpoint, and $\delta_{l}>0$ and $\delta_{u}<1$ denote the infimum and supremum of the values of the function, respectively. Then, we set
    \begin{equation*}
        p^i_{\text{No}}(\pi[k]) = 1 - p^i_{\text{Yes}}(\pi[k]).
    \end{equation*}
    \item As such, when a vehicle $i$ enters the \emph{area access entry gate} at time step $k$, its dynamics are described by
    \begin{align}
        x^i[k+1] & = b^i, \label{eq:update-x}\\
        y^i[k] & = x^i[k], \label{eq:update-y}
    \end{align}
    {where $b^i$ is a random variable selected from the set $\{0,1\}$ according to $\mathbb{P}(b^i=0)=p^{i}_\text{No}(\pi[k])$ and $\mathbb{P}(b^i=1)=p^{i}_\text{Yes}(\pi[k])$}.
    \item Let $k=k+1$. The flow of vehicles travelling on the road of interest at time step $k$ is measured, producing the signal $\hat{y}[k]$.
    \item The signal $\hat{y}[k]$ is compared to $r$, resulting in an update to $e[k]$, and the algorithms continues in an iterative manner.
\end{enumerate}

\subsection{Choice of Probability Functions}
\label{sec:system_design}

The long-term predictability of the evolution of our algorithm described in Section \ref{Algorithm_Description} is guaranteed on average thanks to \cite[Theorem 12]{fioravanti_ergodic_2019}. Specifically, our imposed assumptions on the controller, filter, vehicle dynamics, and probability functions, ensure that the feedback system will converge in distribution to a \emph{unique invariant measure}. That is to say, regardless of any initial condition, as our system evolves over time, its state distribution will approach a fixed, unchanging distribution. Indeed, the assumption of stability on the controller is important; for instance, in \cite{fioravanti_ergodic_2019} it is demonstrated that a classic PI controller can introduce undesirable behaviours due to the pole at $z=1$. Therefore, in our system design, we consider a lag controller of the form
\begin{equation*}
    \pi[k] = \beta\pi[k - 1] + \kappa(e[k] - \alpha e[k - 1]),
\end{equation*}
where $\alpha$, $\beta \neq 1$ and $\kappa$ are tunable parameters.

\begin{figure}[h]
    \centering
\includegraphics[width=\columnwidth]{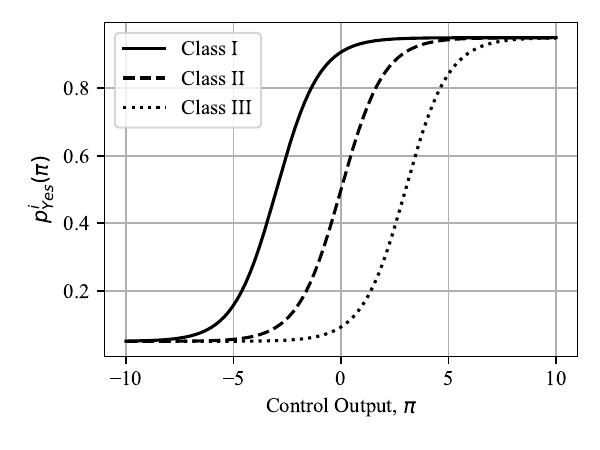}
    \caption{Probability functions used to shape the access probability for different vehicle classes as functions of traffic levels, i.e., of the control output $\pi$.}
    \label{fig:probability_functions}
\end{figure}

Moreover, our framework allows us to assign different logistic functions to different classes of vehicles based on their permitted maximum momentum, $\rho^i_{max}$ computed as in Equation (\ref{rho_max}), as they approach the \emph{road entry access gate}. Thus, we can influence the proportions of vehicles granted access by our system, to the road of interest, in the long-run, according to vehicle class; we can do this, while still allowing for ``fairness'' in the sense that vehicle types with a lower probability of being granted access will nonetheless, on occasion, be permitted through (e.g., also SUVs can be allowed to access the restricted area when traffic conditions are convenient). As an example, we consider the following three classes of vehicles:
\begin{enumerate}[leftmargin=20mm]
    \item[Class I:] $\rho^i_{max} \leq 10^4 \ \mathrm{kg}\cdot \mathrm{m} \cdot \mathrm{s}^{-1}$ (e.g., motorbikes of 300 kg or less, travelling at 120 km/h);
    \item[Class II:] $10^4 \ \mathrm{kg}\cdot \mathrm{m} \cdot \mathrm{s}^{-1} < \rho^i_{max} \leq 3.3 \cdot 10^5 \ \mathrm{kg}\cdot \mathrm{m} \cdot \mathrm{s}^{-1}$ (e.g., conventional cars);
    \item[Class III:] $\rho^i_{max} > 3.3 \cdot 10^5 \ \mathrm{kg}\cdot \mathrm{m} \cdot \mathrm{s}^{-1}$ (e.g., large trucks of 20,000 kg or more, travelling at 60 km/h).
\end{enumerate}
We associate each of these three classes with a different logistic function, as illustrated in Figure \ref{fig:probability_functions}. Since the logistic functions of vehicles of Class I are to the left of those of vehicles of the other classes, then the vehicles of Class I are prioritized (i.e., their probability to be admitted in the restricted area is larger than that of the other vehicles).\\

\textbf{Remark: }For simplicity, we only consider three classes of vehicles here, but more --- and different --- classes may be obviously considered. By appropriately designing and tuning the probability functions of Figure \ref{fig:probability_functions}, it is possible to obtain a different desired target distribution of different classes of vehicles in the restricted area, under different traffic conditions.

\subsection{Access Control Scenario: Case Study}

In this section, we provide a practical evaluation of the feedback based access control strategy using the microscopic traffic simulator SUMO \cite{lopez_microscopic_2018}. All aspects of the simulation are now provided in detail.

\subsubsection{Configuration of the road network}
We consider a simple road network consisting of three edges (corresponding to the one illustrated in Figure \ref{fig:road_access_scenario}) and denote the approaching road as \textit{`approach'}, the road to the restricted area as \textit{`main'} and the alternative road as \textit{`alt'}. Speed limits for \textit{approach}, \textit{main}, and \textit{alt} are set at ${50,100,100}$ km/h respectively. A sensor gate is placed on the \textit{approach}, 30 m before the junction.

\subsubsection{Classes of vehicles}
We consider a heterogenous flow of three vehicle types: motorcycles (200 kg, $v_{max}$ = 144 km/h), cars (2 t,  $v_{max}$ = 144 km/h), trucks (20 t, $v_{max}$ = 72 km/h). At each simulation step, a new vehicle appears at the beginning of the \emph{approach} road with probability $p_{\text{app}}$, with equal probability of belonging to one of the three classes, yielding equal expected appearance rates per class. Upon crossing the sensor gate, vehicles are probabilistically admitted to \emph{main} (Eq. \ref{eq:update-x}) or rerouted to \emph{alt}, and leave the network at the downstream node.

\subsubsection{Controller Design}
We tune the parameters of the lag controller as $\alpha=0.9$, $\beta=0.99$, $\kappa=1$ ($\alpha$ and $\beta$ are usually chosen close to 1, with $\beta > \alpha$, to achieve an appropriate low-frequency gain boost, while $\kappa$ is usually selected to control the overall scaling and cross-over \cite{ogata1995discrete}). The desired access rate is set as $r=9$ veh/min. 

\subsubsection{Filter Design}
The plant measurement, $\hat{y}[k]$ (also in veh/min) is obtained by passing the output, $y[k]$ through a causal $H=20$-sample moving average filter $\mathcal{F}$:
\begin{equation*}
\hat{y}[k] = \bigl(\mathcal F\{y\}\bigr)[k] =\frac{60}{H}\sum_{j=k-H+1}^{k} y[j], \qquad H=20.
\end{equation*}

\subsubsection{Parameters of the logistic probability functions}
\label{Logistic_Function_Parameters}

Logistic‐function parameters $(\delta_l,\delta_u,\lambda,\pi_0)$ are tuned separately for the three momentum classes specified in the preceding section and illustrated in Fig. \ref{fig:probability_functions}:

\begin{equation*}
    (\delta^i_{l},\delta^i_{u},\lambda^i,\pi^i_{0}) =
\begin{cases}
(0.05,\;0.9,\;1,\;-3), &\text{Class (I)} \\
(0.05,\;0.9,\;1,\; 0), &\text{Class (II)} \\
(0.05,\;0.9,\;1,\; 3), &\text{Class (III)}.
\end{cases}
\end{equation*}

\subsubsection{Simulation Results}
\begin{figure}[t]
    \centering
     \includegraphics[width=\linewidth]{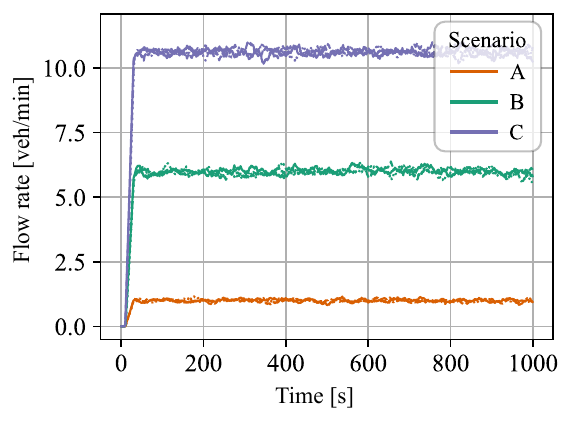}
        \caption{Mean vehicle arrival rates [veh/min] per scenario and class over time.}
        \label{fig:arrival_rates}
\end{figure}

We consider three different scenarios corresponding to three different levels of traffic, namely, a low-traffic case (A), medium case (B) and high-traffic case (C), corresponding to average flows of 1, 6 and 11 vehicles per vehicle class per minute respectively, as depicted in Figure \ref{fig:arrival_rates}. For each traffic demand scenario, we run 1000 independent simulations, where each run lasts 1000 seconds. 
\begin{figure}[t]
    \includegraphics[width=\linewidth]{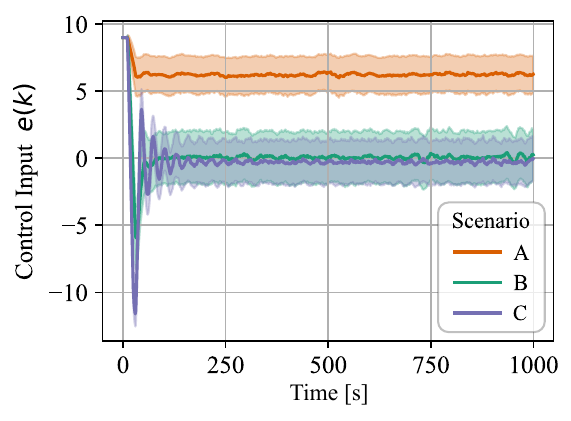}
    \caption{Mean control input $e(k)$. Shaded area represents half standard deviation, $\sigma/2$.}
    \label{fig:control_input}
\end{figure}%
Figure \ref{fig:control_input} shows that in Scenarios B and C the access control system manages to guarantee a flow rate very close to the desired one, and the average error with respect to the target flow is close to zero. Conversely, in scenario A, there is a constant error different to zero: this implies that the restricted area is underutilized, with a flow lower than the target one, but this is due to the fact that only few vehicles ask to access the area (e.g., as in a night scenario). Thus, even allowing all vehicles to access the restricted area, the flow rate of vehicles is smaller than the reference value.

\begin{figure}
    \centering
    \includegraphics[width=\linewidth]{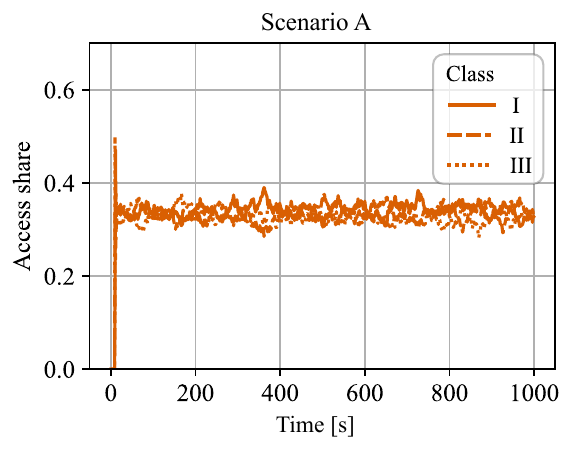}
    \caption{Mean access share of classes I–III in Scenario A over time.}
    \label{fig:access_share_a}
\end{figure}

\begin{figure}
    \centering
    \includegraphics[width=\linewidth]{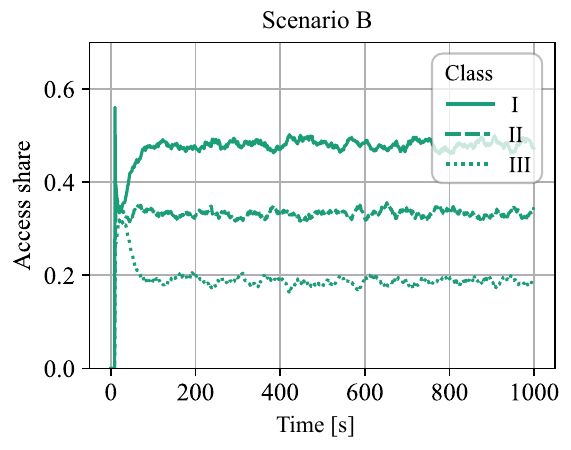}
    \caption{Mean access share of classes I–III in Scenario B over time.}
    \label{fig:access_share_b}
\end{figure}

\begin{figure}
    \centering
    \includegraphics[width=\linewidth]{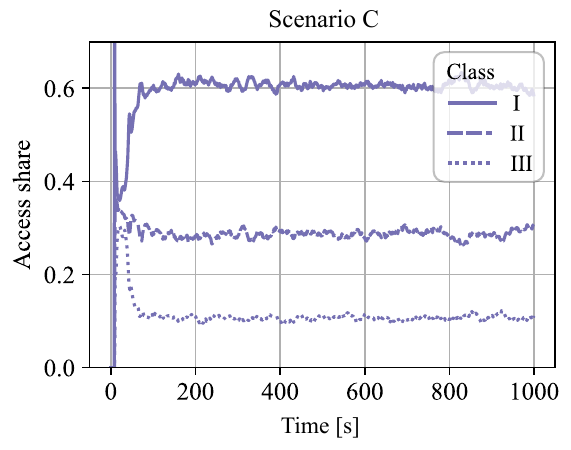}
    \caption{Mean access share of classes I–III in Scenario C over time.}
    \label{fig:access_share_c}
\end{figure}

Figures \ref{fig:access_share_a}-\ref{fig:access_share_c} show how access to the \textit{main} road is distributed across vehicle classes in scenarios A-C. In all scenarios, the distribution converges to a steady state after approximately 100 seconds. In scenario A (low traffic demand), very few vehicles ask for access to the shared road \textit{main}, and thus, practically all requests are approved. Accordingly, Figure \ref{fig:access_share_a} shows that the composition of the flow remains the same as the flow of vehicles asking for access in Figure \ref{fig:arrival_rates} (i.e., approximately the same number of vehicles for all classes). When the level of traffic increases, some of the vehicles' requests for access are not approved, and especially heavy vehicles are excluded from the restricted area to decrease heterogeneity of traffic and prioritize lighter vehicles (see Figure \ref{fig:access_share_b}). This effect is even more noticeable in Figure \ref{fig:access_share_c}, where only 10\% of the heavy vehicles are guaranteed access, vs., 60\% of lightest vehicles.\\

\textbf{Remark: }By appropriately tuning the parameters of the logistic probability functions (as in Section \ref{Logistic_Function_Parameters}) it is possible either decrease or increase the differences in the acceptance of the different classes of vehicles in the restricted area, under different traffic conditions.

\section{Momentum-based Speed Control}
\label{sec:Speed_Control}

\subsection{Problem Formulation}
\label{Momentum_Introduction}
Section \ref{sec:Access_Control} addressed the access control to a restricted area based on momentum, and we showed that in this way it is already possible to reduce the heterogeneity of traffic flow, for example by prioritising access for lighter vehicle classes. In this section we further control the speed of the vehicles that have been admitted inside the area. While this problem has been already addressed in the literature \cite{ma2023, li2023, asadi2021}, for instance to reduce pollution, here we address it from a different and entirely novel perspective, as we explicitly take into account the momentum-derived $\Delta v$ metric as it is recognized as the primary factor for influencing the severity of collision (see discussion in Section \ref{Momentum_metrics}). 
In particular, we shall compute the optimal speed for each vehicle that is feasible with classic constraints (vehicular or road speed limits) and at the same time guarantees that the value of $\Delta v$ in the event of rear-end collisions is not too dangerous. Since in principle different speeds are recommended to different vehicles (e.g., according to their masses), then the proposed methodology can be applied when vehicles are driving along roads with more than one lane (i.e., in order to make overtaking possible).

The classic speed control problem is usually formulated as:

\begin{equation}
\label{Optimal_Speed_Problem}
\left\{
\begin{array}{l}
\max\limits_{v_i} \sum\limits_{i=1}^n v_i\\ 
\underline{v_i}\leq v_i \leq \overline{v_i}
\end{array}
\right..
\end{equation}
According to the optimization problem (\ref{Optimal_Speed_Problem}), vehicles try to maximize their speeds $v_i$ --- to minimize travel time --- provided they are within permitted limits. Here, the upper bound $\overline{v_i}$ either corresponds to the speed limit of the road, or to the maximum allowed speed of the vehicle. Some roads (e.g., highways) also indicate a minimum allowed speed $\underline{v_i}$, or if not, the latter term may be simply posed equal to zero. The obvious solution is that vehicles simply travel at speed $\overline{v_i}$.

We now explain how a further constraint may be added to take into account safety along the road in terms of momentum (i.e., velocity should be automatically decreased when approaching a slower/lighter vehicle). 

For this purpose, we recall that $\Delta v_i$, i.e., the absolute speed change for vehicle i, is given as
\begin{equation}
    \Delta v_i=|v_i^{b} - v_i^{a}|,
    \label{eq:Delta_v}
\end{equation}
where $v_i^{b}$ and $v_i^{a}$ denote the respective velocities of vehicle $i$ before and after a potential collision. Following \cite{evans_driver_1994}, let us assume a perfectly inelastic, one-dimensional rear-end collision between two vehicles, $i$ and $j$, where $i$ is closing on $j$. Here, both vehicles travel at the same velocity post collision, and based on the law of conservation of momentum, the momentum of both vehicles post collision is equal to the sum of their momenta prior to collision. Thus, the shared velocity post collision, $v_{i}^{a} = v_{j}^{a}$ can be estimated as

\begin{equation}
    v_{i}^{a}= v_j^{a}=\frac{m_iv_i^{b}+m_jv_j^{b}}{m_i+m_j}.
\end{equation}

Thus, by substituting into \ref{eq:Delta_v}, we obtain that the estimated speed change for vehicles $i$ and $j$ resulting from a potential collision is

\begin{equation}
    \Delta v_i = \frac{m_j}{m_i+m_j}|v_i^{\text{b}} - v_j^{\text{b}}|,
    \label{self_harm}
\end{equation}
and
\begin{equation}
    \Delta v_j = \frac{m_i}{m_i+m_j}|v_i^{\text{b}} - v_j^{\text{b}}|.
    \label{harm_to_others}
\end{equation}

Since our objective is to limit the possible values of $\Delta v$ of close vehicles, for safety reasons, we add two further constraints 
\begin{equation}
\label{eq:delta_v_constraints}
\left\{
\begin{array}{l}
\Delta v_i \leq \overline{\Delta v_i}, \qquad \forall i, \forall j\in S_i\\ 
\\
\Delta v_j \leq \overline{\Delta v_j}, \qquad \forall i, \forall j\in S_i
\end{array}
\right.,
\end{equation}
where $\overline{\Delta v_i}$ and $\overline{\Delta v_j}$ are two prescribed safety thresholds that should not be exceeded, while the set $S_i$ includes all vehicles preceding vehicle $i$ at a lesser speed along the same road (in our simulation, up to a distance of 300 meters).
\\

\textbf{Remark: }In constraints (\ref{eq:delta_v_constraints}), the threshold $\overline{\Delta v_i}$ refers to the maximum difference in speed that, in case of collision with vehicle $j$, would be experienced by the hitting vehicle $i$ (and thus, is related to the harm that the driver of vehicle $i$ may experience); likewise, the threshold $\overline{\Delta v_j}$ denotes the maximum difference in speed that, in case of collision with with vehicle $j$, would be experienced by the hit vehicle $j$ (and thus, is related to the harm that vehicle $i$ may cause to the drivers of the other vehicles). It is convenient to have two different thresholds because it allows one to differentiate between the risk of the striking vehicle and the risk of the struck vehicle. Most notably, it also allows to address the asymmetric risk in a front-rear collision, where the passengers of the hitting vehicle have a higher probability of injury, that should be computed according to the curves of a frontal collision (\ref{fig:injury_probability_frontal}), rather than those of the hit vehicle, whose probability of injury is computed according to rear-end curves (\ref{fig:injury_probability_rear-end}). For instance, in our simulations we select $\overline{\Delta v_i} = 23.19 km/h$ and $\overline{\Delta v_j} = 30.86 km/h$, which correspond to a 1\% probability of a
MAIS3+ injury for both the hitting and the hit vehicle.\\

\textbf{Example: }
As an example, consider an overtake scenario, in which a 200 kg motorcyclist passes a 20000 kg heavy goods vehicle (HGV) with a closing speed of 36 km/h. In this scenario, the motorcyclist poses a very small $\Delta v_j = 0.1$ m/s risk to the HGV driver, while exposing itself to a larger $\Delta v_i = 9.9$ m/s risk. Vice versa, a HGV overtaking a motorcyclist at the same closing speed poses a large $\Delta v_j = 9.9$ m/s risk to the motorcyclist, and a small $\Delta v_i = 0.1$ m/s risk to itself.

\subsection{Computation of Vehicular Speed Reference in Momentum-based Speed Regulation}

The constrained optimization problem (Problem (\ref{Optimal_Speed_Problem}), with momentum-based constraints (\ref{eq:delta_v_constraints})) can be solved in a single step by choosing the maximum from a set of feasible velocities for each vehicle $i$. The feasible set for $v_i$ is given by intersecting the general speed limits, $\underline{v_i}$ and $\overline{v_i}$, and the upper bound on $v_i$ resulting from the $\Delta v$ constraints (Eq. \ref{eq:delta_v_constraints}). Note that the constraints on $\Delta v_i$ and $\Delta v_j$ may be conveniently converted into practical constraints to the closing speed, and thus to speed reference signals for the vehicles,
\begin{align}
    \Delta v_i = \frac{m_j}{m_i+m_j}|v_i^{\text{b}} - v_j^{\text{b}}|\leq {\overline{\Delta v_i}}\\ \implies |v_i^b-v_j^b| \leq \frac{m_i+m_j}{m_j}\overline{\Delta v_i}, 
\end{align}
and
\begin{align}
    \Delta v_j = \frac{m_i}{m_i+m_j}|(v_i^{\text{b}} - v_j^{\text{b}})|\leq \overline{\Delta v_j}\\ \implies |v_i-v_j| \leq \frac{m_i+m_j}{m_i}\overline{\Delta v_j},
\end{align}
with the tighter of the two being the binding one. Thus, let $\overline{r}$ be the maximum allowable closing speed, which is given by
\begin{equation}
    \overline{r} := min \left( \frac{m_i+m_j}{m_j}\overline{\Delta v_i} \,,\frac{m_i+m_j}{m_i}\overline{\Delta v_j} \right),
\end{equation}
then, the feasible set for vehicle i's velocity, $v_i$ resulting from the constraints of  (\ref{Optimal_Speed_Problem}) and  (\ref{eq:delta_v_constraints}) is
\begin{equation}
    v_i \in \left[ \underline{v_i}, min(\overline{v_i}, v_j+\overline{r}) \right],
\end{equation}
and since our objective is to maximize throughput, the solution to the optimization is the maximum value of the feasible set,
\begin{equation}
    v_i^* = max \left[ \underline{v_i}, min(\overline{v_i}, v_j+\overline{r}) \right].
    \label{eq:v_i^o_solution}
\end{equation}
represents the reference speed for the $i$th vehicle. Equation (\ref{eq:v_i^o_solution}) may be interpreted as: a vehicle should travel at the maximum allowed speed $\overline{v_i}$, unless causing a significant risk to itself or another exposed vehicle (slower and possible lighter) in which case the speed should be reduced to $v_j+\overline{r}$; and in any case, the minimum speed $\underline{v_i}$ of the road, if existing, should be respected.

\subsection{Speed Control Scenario: Case Study}
We now illustrate examples of the proposed momentum-based speed advisory system in the form of simple overtake manoeuvres. We assume that the controlled area is a two-lane fast highway ($\overline{v} = 130$ km/h, $\underline{v} = 60$ km/h), and we run simulations using the open-source Simulation for Urban Mobility (SUMO) software package \cite{lopez_microscopic_2018}.\\

\subsubsection{Fast vehicle}
We consider three different possibilities of fast vehicle $i$ that will overtake vehicle $j$ during the simulation: a motorcycle ($m=300$~kg), a passenger vehicle (PV, $m=2500$~kg), and a heavy goods vehicle (HGV, $m=20000$~kg). \\

\subsubsection{Leading (slow) vehicle}
We consider three overtaking scenarios, which differ in the leading vehicle $j$ that will be overtaken by vehicle $i$:
\begin{enumerate}[label=(\Alph*)]\itemsep0.2em
    \item \textbf{Slow HGV:} The leading vehicle $j$ is a HGV ($m=20000$~kg) travelling at $60~km/h$;
    \item \textbf{Slow Motorcycle:} The leading vehicle  $j$ is a motorcycle ($m=300$~kg) travelling at $60~km/h$.
    \item \textbf{Standing PV:} The leading vehicle $j$ is a stationary PV ($m=2500$~kg), i.e., a stranded vehicle along the highway;
\end{enumerate}
Every second, speeds reference signals are updated according to Equation (\ref{eq:v_i^o_solution}).
As mentioned, bounds $\overline{\Delta v_i}$ and $\overline{\Delta v_j}$ are set equal to $23.19$ km/h and $30.86$ km/h respectively, to reflect a 1\% probability of a MAIS3+ injury based on the injury probability curves (fig. \ref{fig:injury_probability_frontal}, \ref{fig:injury_probability_rear-end}) for both vehicles in case of a collision. \\  

\subsubsection{Results}
\begin{figure*}
    \centering
    \includegraphics[width=\linewidth]{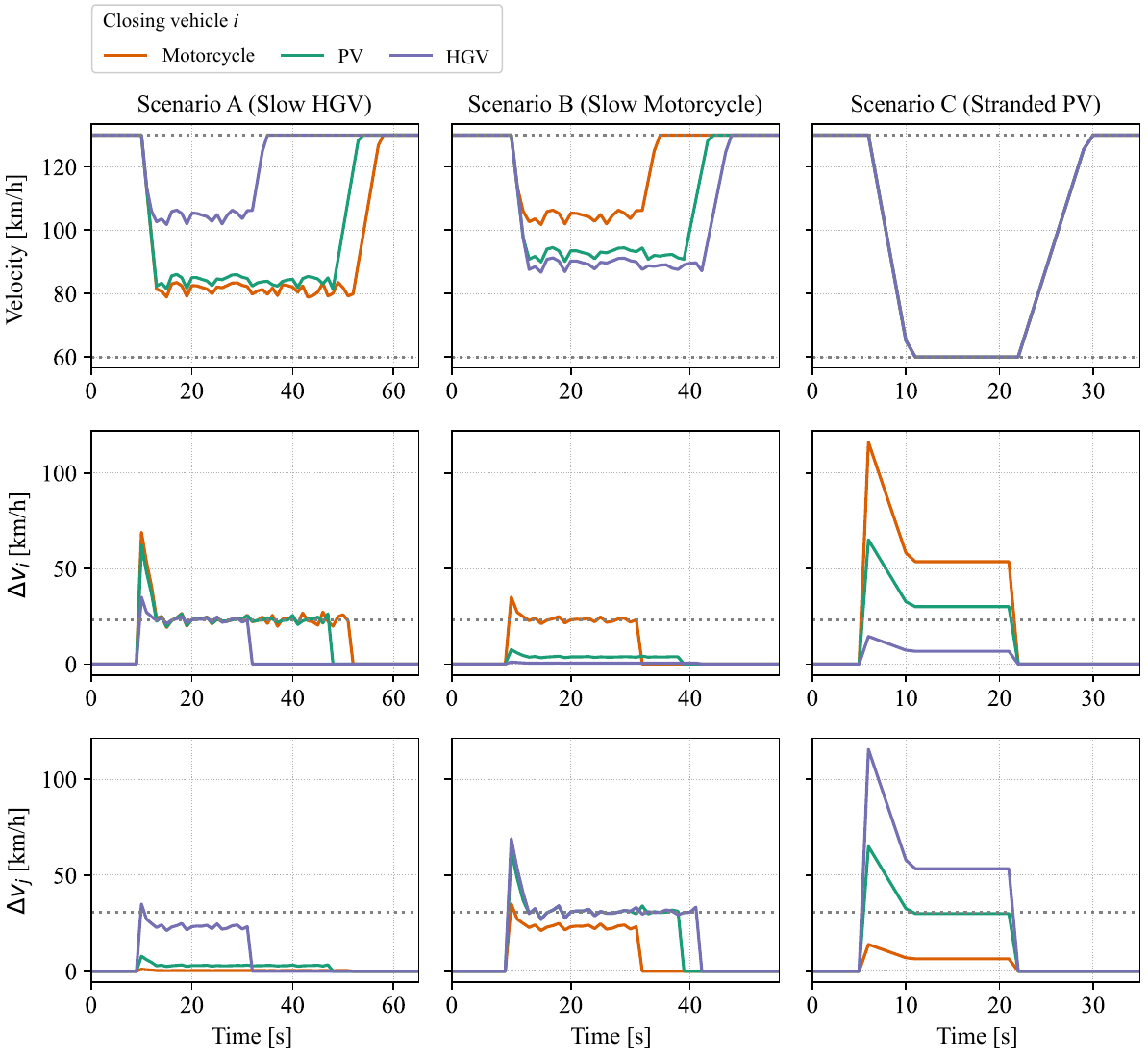}
    \caption{Results of the three overtaking scenarios (columns A–C). The first row shows the velocity profiles of the controlled vehicles when approaching the slower or stranded vehicle. The second and third rows show the safety indicators $\Delta v_i$ and $\Delta v_j$ relative to the safety threshold. Scenario A: overtaking a slow HGV; Scenario B: overtaking a slow motorbike; Scenario C: encountering a stranded car with a 60 km/h minimum speed constraint.}
    \label{fig:Delta_v_experiment}
\end{figure*}

Figure \ref{fig:Delta_v_experiment} summarizes the three overtaking manoeuvres of the three different vehicles in the three scenarios.

In scenario A (first column), a slow moving HGV is overtaken. All controlled vehicles slow down significantly as they approach the slow vehicle due to its large momentum (first row). Both safety indicators $\Delta v_i$ and $\Delta v_j$ are below the safety threshold, apart from a short transient when the slow vehicle is first noticed (i.e., when it is 300 metres away), and some time is required to slow down to the new speed reference signal. The most critical safety indicator is the safety for the faster $i$ vehicle (second row), as it would experience a significant reduction in speed in case of impact with the slow and heavy HGV vehicle.
Scenario B (second column) shows results very similar to those of Scenario A. The main difference is that vehicles reduce their speeds by a lower quantity than in the previous scenario, as the slow vehicle has a smaller momentum (a motorbike instead of an HGV). A second difference is that the most critical safety indicator is safety for the light preceding vehicle (third row), rather than for the vehicle itself (second row).

In Scenario C (third column) the vehicles significantly reduce their speeds to avoid exposing the stranded car to a large risk, but the lower speed limit of the highway prevents them from slowing below 60 km/h. Due to this constraint, then the safety momentum margins are not enforced (second and third row). Constraints could be met if slowing down below 60 km/h was allowed.

\section{Momentum-based Access and Speed Control}
\label{Final_Simulation}

In this section we now evaluate how the proposed momentum-based access and speed control strategies -- singularly and combined -- contribute to improve road safety and efficiency. All our results for each investigated scenario are obtained after 100 Monte Carlo simulations, each lasting 1 hour (3600 time steps), to average the stochastic effects of vehicular flows and traffic.

\subsection{Parameters of the Case Study}
The parameters for each individual Monte Carlo run are chosen as follows:
\subsubsection{Restricted Area}
Similarly to the case study depicted in Figure \ref{fig:road_access_scenario}, we assume that the restricted area consists of a 800 m road with three lanes, such that vehicles travelling on the road may overtake one another. The road speed limit is set at 130 km/h, in line with a typical European highway.
\subsubsection{Vehicular Flow}
We assume that a flow of 24 veh/min requests access to the restricted area, and the reference value for the flow of vehicles is $r = 12$ veh/min (so about 50\% of the vehicles will be denied access to the restricted area). The vehicular flow consists of the three categories of vehicles considered so far (HGVs, motorcycles and PVs) with equal probability of belonging to each class. We also assume that - with probability 0.3 - some HGV vehicles have an upper speed limit of 60 km/h. Accordingly, some of the other vehicles will try to take over such slow HGVs. 
\subsubsection{Parameters of the Access and Speed Control Strategies}
\label{Tuning_ABCD}
We consider the following parameters of the logistic function, in order to ergodically achieve a steady-state distribution of vehicles that aims at strongly prioritizing motorcycles, and then PVs, over HGVs, under the considered heavy traffic conditions:
\begin{equation*}
    (\delta^i_{l},\delta^i_{u},\lambda^i,\pi^i_{0}) =
\begin{cases}
(0.90,\;0.09,\;1,\;-3), &\text{Class (I)} \\
(0.3,\;0.69,\;1,\; 0), &\text{Class (II)} \\
(0.01,\;0.98,\;1,\; 3), &\text{Class (III)}.
\end{cases}
\end{equation*}
Similarly to Section \ref{Momentum_Introduction}, we consider $\overline{\Delta v_i} = 23.19$ km/h and $\overline{\Delta v_j} = 30.86$ km/h, which gives rise to a 1\% probability of a MAIS3+ injury in the event of a collision for both the striking and the struck vehicle.

\subsection{Evaluation of Speed and Access Control Strategies}
We now consider four different scenarios where speed and access control strategies are alternatively active or not, as follows.
\begin{table}[h!]
\centering
\caption{Scenarios considered in the speed and access control evaluation.}
\begin{tabular}{|c|c|c|}
\hline
Scenario/Control strategy & Speed control & Access control \\ \hline
Scenario A   & \ding{55}  & \ding{55}  \\ \hline
Scenario B   & \ding{51}  & \ding{55}  \\ \hline
Scenario C   & \ding{55}  & \ding{51}  \\ \hline
Scenario D   & \ding{51}  & \ding{51}  \\ \hline
\end{tabular}

\label{Final_Scenarios}
\end{table}

\begin{figure}[t]
    \centering
     \includegraphics[width=\linewidth]{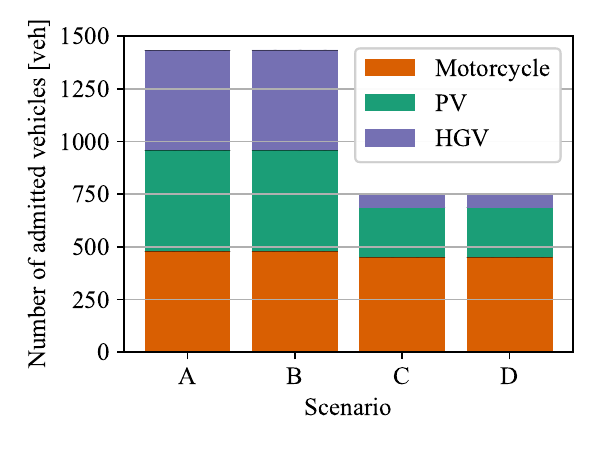}
        \caption{Composition of the vehicular flow in the different scenarios.}
        \label{class_mix_ABCD}
\end{figure}
Figure \ref{class_mix_ABCD} shows the average composition of the vehicles admitted in the restricted area. As can be noticed, when access control is active (Scenarios C and D), only half of the vehicles are admitted (as per the reference value), and light vehicles are more likely to be admitted, thanks to the tuning of the utility functions as described in Section \ref{Tuning_ABCD}. 

\begin{figure}[t]
    \centering
     \includegraphics[width=\linewidth]{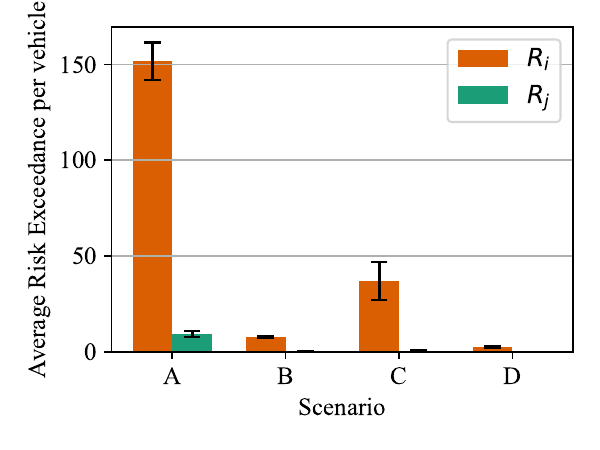}
        \caption{Average risk threshold exceedance in the different scenarios.}
        \label{exceedance_self_vs_other_ABCD}
\end{figure}

Figure \ref{exceedance_self_vs_other_ABCD} summarises the average risk to which each vehicle is exposed during its trip. In particular, we compute the risk of the $i$th vehicle as
\begin{equation}
    \begin{aligned}
    & R_i = \sum_{k=t_\alpha}^{t_\omega} max(\Delta v_i(k) - \overline{\Delta v_i}, 0)\\
    & R_j = \sum_{k=t_\alpha}^{t_\omega} max(\Delta v_j(k) - \overline{\Delta v_j}, 0),
    \end{aligned}
\end{equation}
where $t_\alpha$ is the instant in time when the vehicle starts travelling in the restricted area, and $t_\omega$ is the instant in time when the vehicle leaves the restricted area. 

\begin{figure}[t]
    \centering
     \includegraphics[width=\linewidth]{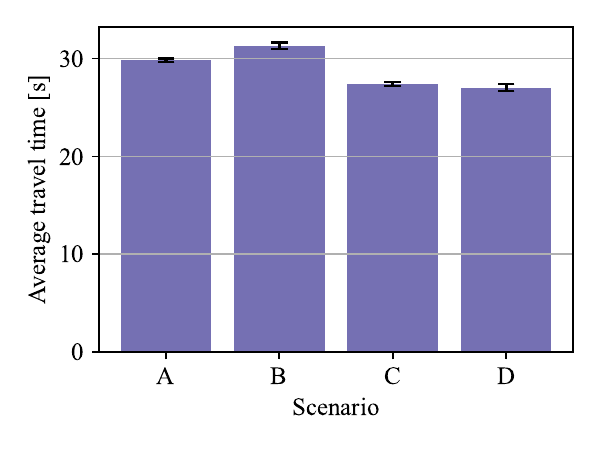}
        \caption{Average travel times on the restricted road.}
        \label{tt_mean_overall_ABCD}
\end{figure}

Finally, Figure \ref{exceedance_self_vs_other_ABCD} shows that travel times are not affected by the proposed strategy, but actually they decrease (as when the number of HGVs decreases, then fewer takeovers are required, and fewer interventions of the speed control system are experienced). Accordingly, this is mainly a consequence of the active access control strategy.

As can be noticed, speed control alone (Scenario B) is effective in drastically improving safety. However, also the access control strategy alone also reduces risk to $1/3$ of the original one without control. This is convenient especially in those areas where speed control strategy cannot be implemented (e.g., because there is no space for overtakes). Finally, the simultaneous application of both speed and access control (Scenario D) is the most effective solution in terms of safety. The few cases when the safety threshold are violated occur as soon as a new vehicle enters the radius of safety (here, 300 metres), before the speed control system manages to reduce the speed into safety margins.

\subsection{Impact of the Volume of Traffic}
\begin{figure}[t]
    \centering
     \includegraphics[width=\linewidth]{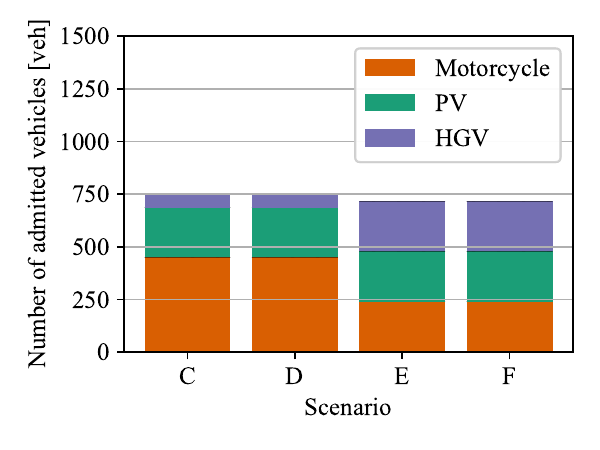}
        \caption{Now four scenarios with the same number of vehicles are compared, and thus the only effect of the access control strategy is thus to change the distribution of the classes of vehicles.}
        \label{class_mix_CDEF}
\end{figure}

Here, we observe that the access control strategy has two net effects (see Figure \ref{class_mix_ABCD}): the first one, is that part of the vehicles are diverted on another road, and fewer vehicles access the restricted area. The second effect, is that the composition of the vehicular flow changes as desired (i.e., through appropriate shaping of the utility functions). In order to clarify the merits of shaping the vehicular flow, and that of simply reducing the number of vehicles, we now consider two further scenarios, that we shall denote as D and E, that are totally equivalent to A and B, but with a reduced number of vehicles, as it is now depicted in Figure \ref{class_mix_CDEF}. 

\begin{figure}[t]
    \centering
     \includegraphics[width=\linewidth]{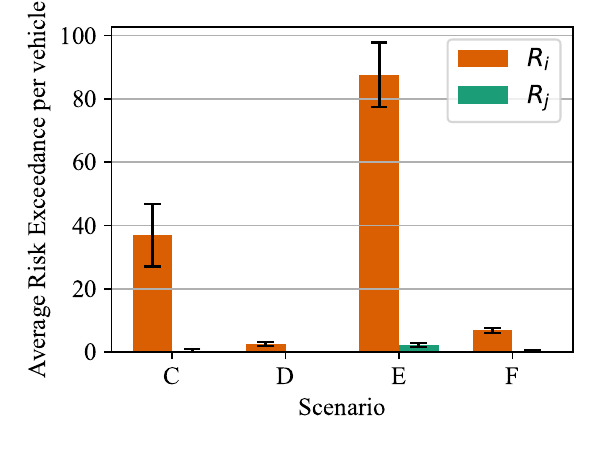}
        \caption{Average risk threshold exceedance in the different scenarios.}
        \label{exceedance_self_vs_other_CDEF}
\end{figure}

\begin{figure}[t]
    \centering
     \includegraphics[width=\linewidth]{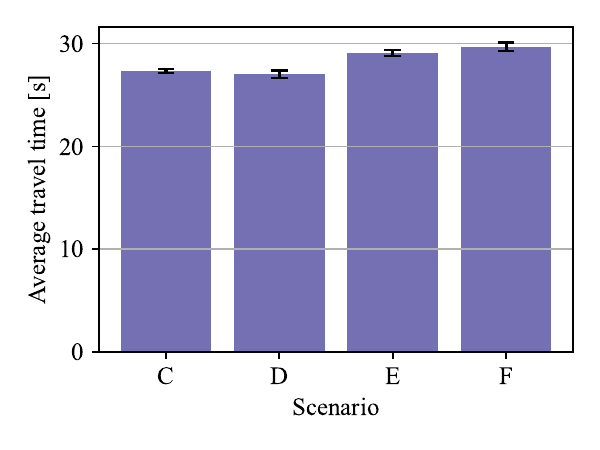}
        \caption{Average travel times on the restricted road.}
        \label{tt_mean_overall_CDEF}
\end{figure}

Accordingly, the only difference between scenarios C and D, and scenarios E and F, is that the vehicular flow is composed of different types of vehicles. 
In particular, Figure \ref{exceedance_self_vs_other_CDEF} shows that even if the total number of vehicles in the restricted area, still the access control strategy alone (C) manages to significantly reduce the risk of the vehicles with respect to the scenario E, where the only difference is given by the composition of the vehicular flow. Further, the utilisation of the speed control system is more effective when access control is active (D) than when access control is not active (F). Finally, shorter travel times are observed again when the access control strategy is active (C and D) as shown in Figure \ref{tt_mean_overall_CDEF}.\\
Thus, the merits of the access control strategy go beyond those of merely reducing the number of vehicles, but are mainly related to the reduction of heterogeneity in the composition of the vehicular flows.

\section{Conclusions}
\label{Conclusions}
Two recent trends in nowadays vehicular networks are the increasing mass of vehicles and the greater heterogeneity of mobility solutions. Increasing vehicle mass is a combined consequence of electrification and consumer preference. To deliver adequate range, electric vehicles require large battery packs, making them significantly heavier than their internal combustion engine equivalents. Simultaneously, larger vehicle types, (e.g., SUVs) are gaining popularity, further contributing to vehicle mass inflation. At the same time, more flexible and lighter mobility solutions are increasingly adopted, such as e-bikes and e-scooters. Such a heterogeneity of means of transportation, and most notably, such heterogeneity of masses, raises significant concerns for the safety of road users, especially as momentum -- and thus the product between mass and speed  -- is recognised as one of the most significant quantities to predict the severity of a road accident.

In this paper we propose two momentum-based speed and access control strategies that, as we show in extensive simulations, manage in practice to reduce the injury risk of drivers. In particular, the access control strategy aims at redesigning the vehicular flows in protected areas, in order to reduce the heterogeneity of traffic in a predictable manner, when this is a concern (e.g., in presence of high volumes of traffic). Conversely, the speed control strategy aims at avoiding too large differences in vehicular speeds, as these could give rise to increased risks of severe accidents. In our work, we show that both strategies are separately able to drastically reduce the risk of road users, while the best results can be obtained when the two strategies are applied simultaneously.

Application of the proposed strategy requires an ad hoc infrastructure to exchange the relevant information, most notably, the masses of the vehicles. As the widespread of vehicular communication networks is increasing, and as more autonomous vehicles are being tested, this appears the right moment to introduce more sophisticated traffic control strategies -- as those suggested in this manuscript -- to improve safety in vehicular networks. Specifically tailored tests and trials of the proposed strategy is indeed our interest as the next step of our work, together with the investigation of the robustness of the proposed strategies in terms of compliance of vehicles and communication delays or failures.

\printbibliography

@article{asadi2021,
  title = {A Systematic Literature Review of Vehicle Speed Assistance in Intelligent Transportation System},
  author = {Asadi, Mehrdad and Fathy, Mahmood and Mahini, Hamidreza and Rahmani, Amir Masoud},
  date = {2021-08},
  journaltitle = {IET Intelligent Transport Systems},
  shortjournal = {IET Intelligent Trans Sys},
  volume = {15},
  number = {8},
  pages = {973--986},
  issn = {1751-956X, 1751-9578},
  doi = {10.1049/itr2.12077},
  url = {https://ietresearch.onlinelibrary.wiley.com/doi/10.1049/itr2.12077},
  urldate = {2025-08-28},
  langid = {english}
}

@article{li2023,
  title = {Developing a {{Dynamic Speed Control System}} for {{Mixed Traffic Flow}} to {{Reduce Collision Risks Near Freeway Bottlenecks}}},
  author = {Li, Ye and Pan, Bin and Chen, Zhibin and Xing, Lu},
  date = {2023-11},
  journaltitle = {IEEE Transactions on Intelligent Transportation Systems},
  shortjournal = {IEEE Trans. Intell. Transport. Syst.},
  volume = {24},
  number = {11},
  pages = {12560--12581},
  issn = {1524-9050, 1558-0016},
  doi = {10.1109/TITS.2023.3287269},
  url = {https://ieeexplore.ieee.org/document/10164644/},
  urldate = {2025-08-28},
  langid = {english}
}

@article{ma2023,
  title = {A {{Review}} of {{Vehicle Speed Control Strategies}}},
  author = {Ma, Changxi and Li, Yuanping and Meng, Wei},
  date = {2023-12},
  journaltitle = {Journal of Intelligent and Connected Vehicles},
  shortjournal = {J. Int. Con. Veh.},
  volume = {6},
  number = {4},
  pages = {190--201},
  issn = {2399-9802},
  doi = {10.26599/JICV.2023.9210010},
  url = {https://ieeexplore.ieee.org/document/10409227/},
  urldate = {2025-08-28},
  langid = {english}
}

@article{perez2010,
  title = {An {{RFID-Based Intelligent Vehicle Speed Controller Using Active Traffic Signals}}},
  author = {P\'erez, Joshu\'e and Seco, Fernando and Milan\'es, Vicente and Jim\'enez, Antonio and D\'iaz, Julio C. and De Pedro, Teresa},
  date = {2010-06-09},
  journaltitle = {Sensors},
  shortjournal = {Sensors},
  volume = {10},
  number = {6},
  pages = {5872--5887},
  issn = {1424-8220},
  doi = {10.3390/s100605872},
  url = {https://www.mdpi.com/1424-8220/10/6/5872},
  urldate = {2025-08-28},
  langid = {english}
}

@article{bhoi2014a,
  title = {Vehicular Communication: A Survey},
  shorttitle = {Vehicular Communication},
  author = {Bhoi, Sourav Kumar and Khilar, Pabitra Mohan},
  date = {2014-09},
  journaltitle = {IET Networks},
  shortjournal = {IET Networks},
  volume = {3},
  number = {3},
  pages = {204--217},
  issn = {2047-4954, 2047-4962},
  doi = {10.1049/iet-net.2013.0065},
  url = {https://ietresearch.onlinelibrary.wiley.com/doi/10.1049/iet-net.2013.0065},
  urldate = {2025-08-28},
  langid = {english}
}

@article{dimitrakopoulos2010,
  title = {Intelligent {{Transportation Systems}}},
  author = {Dimitrakopoulos, George and Demestichas, Panagiotis},
  date = {2010-03},
  journaltitle = {IEEE Vehicular Technology Magazine},
  shortjournal = {IEEE Veh. Technol. Mag.},
  volume = {5},
  number = {1},
  pages = {77--84},
  issn = {1556-6072},
  doi = {10.1109/MVT.2009.935537},
  url = {http://ieeexplore.ieee.org/document/5430544/},
  urldate = {2025-08-28},
  langid = {english}
}

@article{oeschger2020,
  title = {Micromobility and Public Transport Integration: {{The}} Current State of Knowledge},
  shorttitle = {Micromobility and Public Transport Integration},
  author = {Oeschger, Giulia and Carroll, P\'araic and Caulfield, Brian},
  date = {2020-12},
  journaltitle = {Transportation Research Part D: Transport and Environment},
  shortjournal = {Transportation Research Part D: Transport and Environment},
  volume = {89},
  pages = {102628},
  issn = {13619209},
  doi = {10.1016/j.trd.2020.102628},
  url = {https://linkinghub.elsevier.com/retrieve/pii/S1361920920308130},
  urldate = {2025-08-28},
  langid = {english}
}

@techreport{wang_mais0508_2022,
    author = {Wang, Jing-Shiarn},
    title = {{MAIS}(05/08) Injury Probability Curves as Functions of Delta V},
    institution = {National Highway Traffic Safety Administration},
    year = {2022-05}
}

@inproceedings{lopez_microscopic_2018,
	location = {Maui, {HI}},
	title = {Microscopic Traffic Simulation using {SUMO}},
	url = {https://ieeexplore.ieee.org/document/8569938/},
	doi = {10.1109/itsc.2018.8569938},
	eventtitle = {2018 21st International Conference on Intelligent Transportation Systems ({ITSC})},
	pages = {2575--2582},
	booktitle = {2018 21st International Conference on Intelligent Transportation Systems ({ITSC})},
	publisher = {{IEEE}},
	author = {Lopez, Pablo Alvarez and Wiessner, Evamarie and Behrisch, Michael and Bieker-Walz, Laura and Erdmann, Jakob and Flotterod, Yun-Pang and Hilbrich, Robert and Lucken, Leonhard and Rummel, Johannes and Wagner, Peter},
	urldate = {2025-07-17},
	date = {2018-11},
	langid = {english},
}

@article{fioravanti_ergodic_2019,
	title = {On the ergodic control of ensembles},
	volume = {108},
	rights = {https://www.elsevier.com/tdm/userlicense/1.0/},
	issn = {0005-1098},
	url = {https://linkinghub.elsevier.com/retrieve/pii/S0005109819303358},
	doi = {10.1016/j.automatica.2019.06.035},
	pages = {108483},
	journaltitle = {Automatica},
	author = {Fioravanti, André R. and Mareček, Jakub and Shorten, Robert N. and Souza, Matheus and Wirth, Fabian R.},
	urldate = {2025-07-17},
	date = {2019-10},
	langid = {english},
	note = {12 citations (Crossref) [2025-07-17]
Publisher: Elsevier {BV}},
}

@article{noauthor_causal_2001,
	title = {Causal influence of car mass and size on driver fatality risk},
	volume = {91},
    author = {L. Evans},
	issn = {0090-0036, 1541-0048},
	url = {https://ajph.aphapublications.org/doi/full/10.2105/AJPH.91.7.1076},
	doi = {10.2105/AJPH.91.7.1076},
	pages = {1076--1081},
	number = {7},
	journaltitle = {American Journal of Public Health},
	shortjournal = {Am J Public Health},
	urldate = {2025-05-29},
	date = {2001-07-01},
	langid = {english},
	note = {31 citations (Crossref) [2025-05-29]},
}

@book{ogata1995discrete,
  author    = {Katsuhiko Ogata},
  title     = {Discrete-Time Control Systems},
  edition   = {2nd},
  year      = {1995},
  publisher = {Prentice Hall},
  address   = {Upper Saddle River, NJ},
  isbn      = {0-13-034281-5}
}

@article{li_traffic_2012,
	title = {Traffic safety and vehicle choice: quantifying the effects of the ‘arms race’ on American roads},
	volume = {27},
	rights = {http://onlinelibrary.wiley.com/{termsAndConditions}\#vor},
	issn = {0883-7252, 1099-1255},
	url = {https://onlinelibrary.wiley.com/doi/10.1002/jae.1161},
	doi = {10.1002/jae.1161},
	shorttitle = {Traffic safety and vehicle choice},
	pages = {34--62},
	number = {1},
	journaltitle = {Journal of Applied Econometrics},
	shortjournal = {J of Applied Econometrics},
	author = {Li, Shanjun},
	urldate = {2025-05-29},
	date = {2012-01},
	langid = {english},
	note = {40 citations (Crossref) [2025-05-29]},
}

@article{white_arms_2004,
	title = {The “Arms Race” on American Roads: The Effect of Sport Utility Vehicles and Pickup Trucks on Traffic Safety},
	volume = {47},
	issn = {0022-2186, 1537-5285},
	url = {https://www.journals.uchicago.edu/doi/10.1086/422979},
	doi = {10.1086/422979},
	shorttitle = {The “Arms Race” on American Roads},
	pages = {333--355},
	number = {2},
	journaltitle = {The Journal of Law and Economics},
	shortjournal = {The Journal of Law and Economics},
	author = {White, Michelle J.},
	urldate = {2025-05-29},
	date = {2004-10},
	langid = {english},
	note = {100 citations (Crossref) [2025-05-29]},
}

@article{sobhani_kinetic_2011,
	title = {A kinetic energy model of two-vehicle crash injury severity},
	volume = {43},
	rights = {https://www.elsevier.com/tdm/userlicense/1.0/},
	issn = {00014575},
	url = {https://linkinghub.elsevier.com/retrieve/pii/S0001457510003143},
	doi = {10.1016/j.aap.2010.10.021},
	pages = {741--754},
	number = {3},
	journaltitle = {Accident Analysis \& Prevention},
	shortjournal = {Accident Analysis \& Prevention},
	author = {Sobhani, Amir and Young, William and Logan, David and Bahrololoom, Sareh},
	urldate = {2025-05-29},
	date = {2011-05},
	langid = {english},
	note = {65 citations (Crossref) [2025-05-29]},
}

@article{tyndall_pedestrian_2021,
	title = {Pedestrian deaths and large vehicles},
	volume = {26-27},
	rights = {https://www.elsevier.com/tdm/userlicense/1.0/},
	issn = {2212-0122},
	url = {https://linkinghub.elsevier.com/retrieve/pii/S2212012221000241},
	doi = {10.1016/j.ecotra.2021.100219},
	pages = {100219},
	journaltitle = {Economics of Transportation},
	author = {Tyndall, Justin},
	urldate = {2024-09-17},
	date = {2021-06},
	langid = {english},
	note = {19 citations (Crossref) [2025-05-29]
Publisher: Elsevier {BV}},
}

@article{evans_driver_1994,
	title = {Driver injury and fatality risk in two-car crashes versus mass ratio inferred using Newtonian mechanics},
	volume = {26},
	rights = {https://www.elsevier.com/tdm/userlicense/1.0/},
	issn = {00014575},
	url = {https://linkinghub.elsevier.com/retrieve/pii/0001457594900221},
	doi = {10.1016/0001-4575(94)90022-1},
	pages = {609--616},
	number = {5},
	journaltitle = {Accident Analysis \& Prevention},
	shortjournal = {Accident Analysis \& Prevention},
	author = {Evans, Leonard},
	urldate = {2025-05-29},
	date = {1994-10},
	langid = {english},
	note = {97 citations (Crossref) [2025-05-29]},
}

@article{evans_mass_1993,
	title = {Mass ratio and relative driver fatality risk in two-vehicle crashes},
	volume = {25},
	rights = {https://www.elsevier.com/tdm/userlicense/1.0/},
	issn = {00014575},
	url = {https://linkinghub.elsevier.com/retrieve/pii/0001457593900622},
	doi = {10.1016/0001-4575(93)90062-2},
	pages = {213--224},
	number = {2},
	journaltitle = {Accident Analysis \& Prevention},
	shortjournal = {Accident Analysis \& Prevention},
	author = {Evans, Leonard and Frick, Michael C.},
	urldate = {2025-05-29},
	date = {1993-04},
	langid = {english},
	note = {75 citations (Crossref) [2025-05-29]},
}

@article{economist2024,
	title = {Americans’ love affair with big cars is killing them},
	issn = {0013-0613},
    year = {2024},
	url = {https://www.economist.com/interactive/united-states/2024/08/31/americans-love-affair-with-big-cars-is-killing-them},
    author = {{The Economist}},
	journaltitle = {The Economist},
	urldate = {2025-05-12},
}

@techreport{NBERw17170,
 title = {Pounds that Kill: The External Costs of Vehicle Weight},
 author = {Anderson, Michael and Auffhammer, Maximilian},
 institution = {National Bureau of Economic Research},
 type = {Working Paper},
 series = {Working Paper Series},
 number = {17170},
 year = {2011},
 month = {June},
 doi = {10.3386/w17170},
 URL = {http://www.nber.org/papers/w17170},
}

\end{document}